\documentclass[a4paper,10pt]{article}
\usepackage{amssymb}
\usepackage{amstext}
\title{%
   On a Possible Interpretation of Fusion in Stochastic L\"owner Evolution
}
\author{%
  {\normalsize\sc Annekathrin M\"uller-Lohmann\thanks{{\tt 
    anne@itp.uni-hannover.de}}}\\[0.5cm]
  {\normalsize\slshape Institut f\"ur theoretische Physik}\\[-0.1cm]
  {\normalsize\slshape Leibniz Universit\"at Hannover}\\[-0.1cm]
  {\normalsize\slshape Appelstra\ss e 2}\\[-0.1cm]
  {\normalsize\slshape D-30167 Hannover, Germany}\\[-0.1cm]
  {\normalsize\slshape Tel.: +49 (0)511 762-4837, Fax: -3023}
}
\date{\today}
\begin{document}

\maketitle

\begin{abstract}
We suggest how to give a physical interpretation of Stochastic L\"owner Evolution traces approaching a marked point in the upper half plane $\mathbb{H}$. We show that this may be related to the fusion of boundary with bulk fields in Conformal Field Theory by taking a look at the probability of such an event.

\textbf{PACS}: 11.25.Hf, 02.50.Ey

\textbf{Keywords}: Conformal Field Theory, Stochastic L\"owner Evolution
\end{abstract}

\section{Introduction}
Eight years ago, a new method designed to show that the scaling limit of certain two-dimensional lattice models in statistical physics at criticality is conformally invariant has been invented by O.\,Schramm \cite{OdedSchramm}. His so called Stochastic L\"owner Evolutions ({\sc SLE}s) are one parameter families of conformally invariant measures $(g_t)_{t\in\mathbb{R}^+_0}$ growing in time according to the {\sc L\"owner} differential equation. 
Requiring conformal invariance and the {\sc Markov} property, its driving term is fixed to {\sc Brownian} motion $\xi_t = \sqrt{\kappa} B_t$ such that its pre-images form a curve $\gamma_t$ that is a random cluster interface connected to the boundary in a statistical physics model at criticality (for an elaborate introduction and proper definition see e.\,g.\ \cite{Lawler}). Generally it is hoped for that {\sc SLE}s can be connected to Conformal Field Theories ({\sc CFT}s), matching the speed of the {\sc Brownian} motion $\kappa$ to the central charge $c = (3\kappa-8)(6-\kappa)/2\kappa$ of the CFT.

The connection between {\sc SLE} and {\sc CFT} is based on lifting the {\sc SLE} to a formal group whose martingales then correspond to null vectors in the {\sc Verma} module of the {\sc Virasoro} algebra \cite{Bauer:2002tf} corresponding to level two descendants of the {\sc CFT}-fields $\psi_{(1,2)}$ or $\psi_{(2,1)}$. Recently, there have been several attempts to relate more than these fields to objects in {\sc SLE} \cite{Rasmussen:2004} 
and multiple {\sc SLE} \cite{Bauer:2005jt,Cardy:2003,Dubedat:2004-2} via the comparison of differential equations. Additionally, the probabilities of visiting small sections on the real or imaginary axis and points in the upper half plane have been investigated \cite{Bauer:2004,Beffara:2002,Friedrich:2002}. In some cases, they have been identified with {\sc CFT} quantities through similar scaling behavior. In others, the stress-energy tensor \cite{Cardy:2005} is identified or special features of the $Q$-states {\sc Potts} model are discussed \cite{Cardy:2006}.

In this paper we will introduce a new way to look upon the martingale-null vector relationship: we strongly believe that we have to include the behavior of the differential operator acting on the associated primary field, since the whole correspondence has been built up on matching entire differential equations, not only the quantities satisfying them. 

Hence when e.\,g.\ investigating the small distance behavior of correlation functions containing a null field, we have to take into account contributions from operator product expansion of the associated primary field with another primary as well as from the differential operator acting on it. In this work we will show that, indeed, the null vector differential operator in {\sc CFT} contributes to the small distance behavior of the differential equation. Therefore the scaling behavior of the correlation functions we want to compare to objects in {\sc SLE} will be modified. This way we will show how to give a new interpretation of the {\sc SLE} probability of hitting a ball of some given radius with the {\sc L\"owner} trace.

\section{Review of Known SLE and CFT Features}
\subsection{SLE Probability of Hitting a Ball}\label{sec:prob}
Considering the {\sc L\"owner} equation for ${0<\kappa\leq 8}$,
\begin{equation}
 \text{d} g_t = \frac{2 \text{d} t}{g_t - \xi_t}\,,
\end{equation}
with initial condition $g_0 = z$ in the hydrodynamical normalization $g_t(z) = z + \frac{2t}{z} + \mathcal{O}(z^{-2})$ at $ z = \infty$, the probability of an {\sc SLE}$_\kappa$ trace $\gamma_t :\,= \lim_{\delta \downarrow 0}g_t^{-1}(\xi_t + \mathrm{i}\delta)$ intersecting a ball $\mathcal{B}_\epsilon (z_0)$ of radius $\epsilon$ centered at a point $z_0$ in the upper half plane is given by \cite{Beffara:2002}:
\begin{eqnarray}\label{eq:Pepsz0}
\textbf{P}_{\epsilon,z_0} &:\,=& \textbf{P}\left(\gamma_{(0,\infty]} \cap \mathcal{B}_\epsilon (z_0) \neq \emptyset \right)\nonumber\\
&\asymp& \left( \frac{\epsilon}{\mathfrak{Im}(z_0)} \right)^{2 - d_{\gamma_t}} \left( \sin \alpha(z_0) \right)^{8/\kappa-1}\,,
\end{eqnarray}
where $d_{\gamma_t} = \text{max}\{2,1+\kappa/8\}$ ist the {\sc Hausdorff} dimension of the {\sc SLE} path and $\alpha(z_0) = \log z_0/\bar{z}_0$. 

\subsection{SLE Martingales \& BCFT Conserved Quantities}
In the physics community, it is widely believed that martingales in {\sc SLE} should correspond to conserved quantities. As such they should be expressible by expectation values $\mathcal{O}$ of observables $O$ that are furthermore believed to be products of primary fields $O (\{z_l,\bar{z}_l\}) = \prod_{l=1}^n \phi(z_l,\bar{z}_l)$ \cite{Bauer:2005jt}. In boundary {\sc CFT} these $n$ local primary fields are regarded as $2n$ independent chiral fields with $z_{n+l} = \bar{z}_l$ in abuse of notation. The boundary conditions are imposed by taking the limit $z_{n+l} \rightarrow z^*_l$ (with $*$ denoting the complet conjugate) after solving the differential equation for the correlation function. Hence, starting with an {\sc SLE} setting at time $t$, i.\,e.\ we already have an interface $\gamma_t$ grown from $0$ to $w_t$, we can say that
\begin{equation}
 \mathcal{O}_t^{\text{\sc SLE}} \rightarrow \frac{\mathcal{O}_{\mathbb{H}_t}^{\text{\sc bCFT}} }{\left\langle \textbf{1} \right\rangle_{\mathbb{H}_t}^{\text{\sc bCFT}}} = \frac{\left\langle  \psi(\infty), ^{g_t}O (\{z_l\}) \psi(\xi_t) \right\rangle_{\mathbb{H}}^{\text{\sc bCFT}}}{\left\langle  \psi(\infty),  \psi (\xi_t) \right\rangle_{\mathbb{H}}^{\text{\sc bCFT}}}\,,
\end{equation}
As correlation functions in {\sc CFT}, the expectation values of observables can be identified through null vector differential equations. Thus investigating the behavior of such objects in the limit $w_t \rightarrow z_l$, we also have to take into account how the respective differential operators change. This will be done in section \ref{sec:limit}. 

In order to translate the vanishing time derivative of the expectation value of our martingale, i.\,e.\ $\frac{\text{d}}{\text{d}t} \mathbb{E} (O^{\text{SLE}}_t) = 0$, to a {\sc CFT} equation, we have to compute the variation of the observable due to $g_t$. Introducing $z_t^l = g_t(z_l)$ etc., we have
\begin{equation}
\text{d} \left(^{g_t}O (\{z_l\}) \right)= -2 \sum_{l=1}^{2n} \left(  \frac{h_l}{\left(z_t^l - \xi_t\right)^2} - \frac{1}{z_t^l - \xi_t}\partial_{z_t^l}
\right) 
\,^{g_t} O (\{z_l\})\text{d}t\,,\nonumber
\end{equation}
Note that in the antiholomorphic part (a.h.) we have to account for the action of the L\"owner mapping on the lower half plane due to the mirror images method which is a standard way to deal with boundary conditions in {\sc CFT}.

The variation of the boundary field gives us
\begin{equation}
\text{d}(\psi(\xi_t)) = \partial_{\xi_t} \psi (\xi_t) \text{d}\xi_t + \frac{\kappa}{2} \partial^2_{\xi_t} \psi (\xi_t) \text{d}t\,.
\end{equation}
This leads us to the {\sc CFT} level two differential equation:
\begin{equation}\label{eq:diff}
\frac{\text{d}}{\text{d}t} \mathbb{E} \left(O_t^{\text{\sc SLE}} \right) = \mathcal{D}_{-2}^{2n}(\xi_t; \{z_t^l\}) \mathcal{O}_t^{\text{\sc bCFT}} (\{z_t^l\}) = 0\,,
\end{equation}
where we defined
\begin{equation}
\mathcal{D}_{-2}^{2n}(\xi_t; \{z_t^l\});\,=\frac{\kappa}{2}\partial^2_{\xi_t} - 2 \sum_{l=1}^{2n} \left(  \frac{h_l}{\left(z_t^l - \xi_t\right)^2} - \frac{1}{z_t^l - \xi_t}\partial_{z_t^l} 
\right)
\end{equation}
which can be identified as the null vector operators on level two in minimal models of {\sc CFT}.
\subsection{Fusion in CFT and the OPE}
In order to be able to study the small distance behavior of products of fields, we have to introduce the concept of fusion and the operator product expansion ({\sc OPE}) of primary fields in {\sc CFT}.

In minimal {\sc CFT}, the primary fields correspond to heighest weights in the Kac table. The so-called fusion rules tell us which primaries and descendants are involved in the short distance product of two given fields. In principle, if the two fundamental fields $\phi_{(2,1)}$ and $\phi_{(1,2)}$ are present in a theory, all other fields may be generated by consecutive fusion of a suitable number of copies. 

The {\sc OPE} for the chiral fields is given by (e.\,g.\ cf.\ \cite{DiFrancesco:1997nk})
\begin{equation}\label{eq:OPE}
\phi_{(r_0,s_0)} (z) \phi_{(r_1,s_1)}(w) \\
=
\sum_{h_{(r',s')},Y} g_{(r',s')} 
 (z-w)^{|Y|-\mu} \beta_{Y,(r',s')} L_{-Y} \phi_{(r',s')} (w)\,,\nonumber
\end{equation}
where $Y = \{ k_1,k_2,\ldots, k_n\}$, $k_1\geq \ldots \geq k_n$, $|Y| = \sum_{i=1}^n k_i$,  $\mu = h_{(r',s')}-h_{(r_0,s_0)}-h_{(r_1,s_1)}$ and $g_{(r',s')}$ the coefficient of the three-point function of the involved fields. It is only nonzero for weights $h_{(r',s')}$ that appear in the fusion product ``$\times$'' of the generators of the fusion algebra $\phi_{(r_0,s_0)}$ and $\phi_{(r_1,s_1)}$ corresponding to the respective fields. 
For our purposes, it suffices to know that
\begin{eqnarray}
 \phi_{(1,2)} \times \phi_{(1,2)} =  \phi_{(1,1)} +  \phi_{(1,3)}\,,\\
 \phi_{(2,1)} \times \phi_{(2,1)} =  \phi_{(1,1)} +  \phi_{(3,1)}\,.
\end{eqnarray}

\section{New Viewpoints and Interpretation}
\subsection{Limits of the Differential Operators}\label{sec:limit}
As we have already stated above, observables in Stochastic L\"owner Evolution that are also {\sc CFT} observables, are believed to be proportional to $O_t\left( \{z_t^l\}\right) $. These correlation functions are known to satisfy the differential equation (\ref{eq:diff}). 
Now we would like to study the small distance behavior of the differential operators $ \mathcal{D}_{-2}^{2n}(\xi_t; \{z_t^l\})$ acting on the OPE of two fields $\psi_{(r,s)}(\xi_t)$ and $\phi_{(r,s)}(z_j^t)$. 
Hence we have $g_h = 0$ for $h \neq h_{(1,1)}$ or $h_{(r',s')}$ with $r' = 2r-1$ and $s' = 2s-1$
where $(r,s) = (1,2)$ or $(2,1)$.
In the following we will restrict ourselves to the $h_{(r',s')}$ branch since $h_{(1,1)}$ has already been studied in \cite{Bauer:2005jt}. 
Introducing the notation $2z_t = \xi_t + z_t^j$ and $\epsilon \,:= \xi_t - z_t^j$, we can reexpress $\mathcal{D}_{-2}^{2n}(\xi_t; \{z_t^l\})$ in the new coordinates including $(\xi_t,z_t^j) \rightarrow (\epsilon,z_t)$, denoted by $\mathcal{D}_{-2}^{2n}(z_t,\epsilon ; \{z_t^l\}_{l\neq j})$:
\begin{equation}
\frac{\kappa}{2}\left(\partial_\epsilon^2 - 2 \partial_\epsilon \partial_{z_t} + \partial_{z_t}^2\right)-  2 \left( \frac{h_{(r,s)}}{\epsilon^2} - \frac{1}{\epsilon} \partial_\epsilon + \sum_{l\neq j=1}^{2n} \frac{h_l}{(z_t^l - z_t)^2} - \frac{1}{z_t^l -z_t}\partial_{z_t^l} \right)\nonumber\,.
\end{equation}
Letting this act on $\mathcal{O}_t (\{z_t^l\})$ with the OPE (\ref{eq:OPE}) inserted, $\mathcal{O}_t (z_t,\epsilon,\{z_t^l\}_{l \neq j})$, we get only contributions from the $|Y|=1$ term\footnote{The procedure how to obtain the $\epsilon$-dependencies is quite lenghty but can in principle be found in (\cite{DiFrancesco:1997nk} -- Appendix 8A). Heuristically, it can be easiest understood via comparison of the dimension on both sides.}:
\begin{equation}
 \mathcal{D}_{-2}^{2n}(z_t,\epsilon ; \{z_t^l\}_{l\neq j}) \propto \epsilon \mathcal{D}_{-3}^{2n-1}(z_t;\{z_t^l\}_{l\neq j}) \,,
\end{equation}
with the level three differential operator given by
\begin{eqnarray}
\mathcal{D}_{-3}^{2n-1}(z_t;\{z_t^l\}_{l\neq j})&:\,=& 
\frac{\kappa}{2}\partial_{z_t}^3 - 2\left(\sum_{l \neq j=1}^{2n} \frac{h_j}{(z_t^l-z_t)^2}-\frac{1}{z_t^l-z_t}\partial_{z_t^l}\right)\partial_{z_t}\nonumber \\
&\quad&+ h_{(r',s')}\left(\sum_{l \neq j=1}^{2n} \frac{2h_l}{(z_t^l-z_t)^3}-\frac{1}{(z_t^l-z_t)^2}\partial_{z_t^l}\right)\,.
\end{eqnarray}
Thus, the infinitesimal generator of a conformally invariant Markov process depending on $n$ coordinates of which one approaches the coordinate of the driving process becomes dependent on the distance $\epsilon$. In order to compare it to {\sc SLE} objects, we have to combine the $\epsilon$-dependencies in the differential equation:
\begin{eqnarray}\label{eq:rhs}
  \mathcal{D}_{-2}^{2n}(z_t,\epsilon \{z_t^l\}_{l\neq j}) \mathcal{O}_t (z_t,\epsilon,\{z_t^l\}_{l\neq j}) 
&=& \!\!\mathcal{D}_{-3}^{2n-1}(z_t;\{z_t^l\}_{l\neq j}) \epsilon^{1-\mu} \tilde{\mathcal{O}}_t (\{z_t^l\}_{l\neq j})\nonumber\\ &\propto& \epsilon^{1-\mu}\,.
\end{eqnarray}
with $\tilde{\mathcal{O}}_t (z_t,\{z_t^l\}_{l\neq j} = \lim_{\epsilon\rightarrow 0}\epsilon^\mu \mathcal{O}_t(z_t,\epsilon,\{z_t^l\}_{l\neq j})$. 

This is the key point of our paper: contrary to \cite{Bauer:2004}, we get an extra factor of $\epsilon$ which will allow us to interpret the event within the field content of the Kac-table.

In the following interpretation, we will use the formulas above, replacing the dependency on $z_t,z_t^l$ etc.\ with a functional dependency on $z,z^l$ given by $t \mapsto z_t,z_t^l$ as introduced in \cite{Bauer:2005jt} since the differential equations are valid for all times $t$.
\subsection{Putting Things Together}

Now let us define the event $A:\,=\{d=\text{dist}(z_0,\gamma)\,:\, d \leq \epsilon \}$, hence $\textbf{P}(a) =\textbf{P}_{\epsilon,z_j}$. Then we know that $\mathbb{E}(O_t^{\text{\sc SLE}}) = \mathbb{E}(O_t^{\text{\sc SLE}}\cdot 1_A + O_t^{\text{\sc SLE}} \cdot 1_{\bar{A}}) \rightarrow \mathbb{E}(O_t^{\text{\sc SLE}}\cdot 1_A)$ since when we can neglect higher order terms on the {\sc rhs} of (\ref{eq:diff}), we can be sure that event $A$ happens since the distance has to be small. Hence using the definition $\mathbb{E}(\tilde{O}_t^{\text{\sc SLE}}) = \mathbb{E}(O_t^{\text{\sc SLE}} | A) = \frac{1}{\textbf{P}(A)}\int_A O_t^{\text{\sc SLE}} \text{d} \textbf{P} = \frac{1}{\textbf{P}(A)}\int O_t^{\text{\sc SLE}}1_A \text{d} \textbf{P} = \frac{1}{\textbf{P}(A)}\mathbb{E}(O_t^{\text{\sc SLE}}\cdot 1_A)$, we see that:
\begin{equation}
 \frac{\text{d}}{\text{d}t} \mathbb{E}(O_t^{\text{\sc SLE}}) \rightarrow  \frac{\text{d}}{\text{d}t} \mathbb{E}(\tilde{O}_t^{\text{\sc SLE}}) \textbf{P}_{\epsilon,z_j} \propto \epsilon^{2-\text{d}_{\gamma_t}}\,,
\end{equation}
according to section \ref{sec:prob}. Comparing this to the {\sc r.h.s.} (\ref{eq:rhs}) we observe $1-\mu = 2-\text{d}_{K_t}$ for an appropriate indentification:
\begin{itemize}
 \item for $0 < \kappa \leq 4$: an {\sc SLE} trace that corresponds to an $(2,1)$-type boundary {\sc CFT} field approaches a bulk field of the same weight located at the point $z_j$. Here we have $1-\mu = 1-\frac{\kappa}{8} = 2-\text{d}_{K_t} = 2-\text{d}_{\gamma_t}$ since $h_{(2,1)} = \frac{3\kappa-8}{16}$ and $h_{(3,1)} = \frac{\kappa-2}{2}$.
 \item for $4 < \kappa \leq 8$: an {\sc SLE} hull locally described by an {\sc SLE} trace with speed $16/\kappa$ that corresponds to an $(1,2)$-type boundary {\sc CFT} field approaches a bulk field of the same weight located at the point $z_j$. Here we have $1-\mu = 1-\frac{2}{\kappa}= 2-\text{d}_{K_t} = 2-\text{d}_{\gamma_t}(\frac{16}{\kappa})$ since $h_{(1,2)} = \frac{6-\kappa}{2}$ and $h_{(1,3)} = \frac{8-\kappa}{\kappa}$.
\end{itemize}
Note that the replacement $d_{K_t}(\kappa) = d_{\gamma_t}(16/\kappa)$ can only be made locally since the duality conjecture for {\sc SLE}s hints at local properties of the trace and the hull \cite{Dubedat:2005}.
\section{Discussion, Conclusion and Perspective}
\subsection{Discussion}
Despite our nice result, some problems remain unsolved. The full formula (\ref{eq:Pepsz0}) contains an angular dependency on $\alpha_0 := \log (z_j / \bar{z}_j)$ and an additional dependency on $\mathfrak{Im}(z_j)$. 
These factors have not yet been related to {\sc CFT} quantities yet. A naive first approach for the angular term can be stated for $n=1$ bulk fields: first transfer back to $\mathbb{H}_t$, 
then use global conformal invariance to arrive at a point where the correlation function is proportional to $|z_j|^{-h_{(1,3)}} = \mathfrak{Im}(z_j)^{-h_{(1,3)}} \sin \alpha_0^{h_{(1,3)}}$. This has in fact already been tried using a field of weight $h_{(0,1)}$ formally of the extended Kac-table \cite{Bauer:2004}. However, this would take us out of the minimal model {\sc CFT} framework and, in this special case, does not provide us with the correct expontent $h_{(1,3)}$ for the angular dependency\footnote{Particularly, taking the fields on the boundary of the Kac-table into account usually leads to logarithmic {\sc CFT}s.
}. 
Luckily this procedure works out fine for us, since our ansatz provides us with an $h_{(1,3)}$ field for $4 < \kappa < 8$, which is the phase where the self-touching {\sc SLE} trace is supposed to come near the boundary again where this factor is dominant. Note that since this term arises from taking the initial condition into account, it has to rely on global properties of the {\sc SLE} curve and hence we do not have to replace $\kappa$ by $16/\kappa$ here. However we have to admit that this formula has not been proven for {\sc SLE} hulls yet and as such our statements for $\kappa > 4$ should be taken as a basis for discussion and not final results already.


\subsection{Conclusion and Perspective}
In this paper we found a {\sc CFT} interpretation for {\sc SLE} exponents that occur in the probability of an {\sc SLE} to come close to a point in the upper half plane. Approaching a bulk point exhibits a behavior similar to the fusion of a bulk Kac-level two field with a boundary field of the same weight. For self-touching {\sc SLE}s with $4 < \kappa < 8$, we proposed an idea how to get the correct angular dependency from fusion of a boundary $h_{(1,2)}$ with a bulk $h_{(1,2)}$ field near the boundary.

We hope to be able to extend this to the fusion of the boundary fields occuring in multiple {\sc SLE} \cite{Bauer:2005jt,Graham:2005,Dubedat:2004-2,Cardy:2003} although the {\sc SLE} probabilities have only been derived for the single {\sc SLE} case yet. Additionally we will try to resolve the difficulties that remain with the dependency on the imaginary part.
\subsection*{Acknowledgements}
I would like to thank Kalle Kyt\"ol\"a and Greg Lawler for really helpful discussions at the IPAM Random Shapes Workshop at UCLA this spring; the IPAM staff and faculty, especially Peter Jones and Christian Ratsch, for its organisation and hospitality and Michael Flohr for further advice. 

This research was funded by the Friedrich-Naumann-Stiftung.
\small
\bibliography{Bibliography}
\end{document}